\documentclass[conference]{IEEEtran}
\usepackage[hyphens]{url}
\usepackage{hyperref}
\hypersetup{colorlinks,allcolors=blue}

\usepackage[backend=bibtex,style=ieee,citestyle=ieee-comp]{biblatex}
\IEEEoverridecommandlockouts

\usepackage{graphicx} 
\usepackage{xcolor} 
\usepackage{minted} 
\usepackage{doi}
\usepackage{orcidlink}
\def\BibTeX{{\rm B\kern-.05em{\sc i\kern-.025em b}\kern-.08em
    T\kern-.1667em\lower.7ex\hbox{E}\kern-.125emX}}

\begin{document}

\title{Dark Path: An Analysis of the Belt \& Road Initiative in El Salvador
\thanks{Disclaimer: The views expressed are those of the authors and do not reflect any official policy or position of the U.S. Department of Defense (War) or the U.S. Government. References to external sites do not constitute endorsement of the same. Cleared for release on 19 MAY 2026 (DOPSR 26-P-0621).}}

\author{\IEEEauthorblockN{Adam Dorian Wong\,\orcidlink{0009-0000-1832-6859}, David Kenley\,\orcidlink{0000-0001-7780-7644}}
\IEEEauthorblockA{\textit{The Beacom College of Computer \& Cyber Sciences} \\
\textit{Dakota State University}\\
Madison, SD, USA \\
adam.wong@trojans.dsu.edu, david.kenley@dsu.edu}}

\maketitle

\begin{abstract} 
The Belt \& Road Initiative (BRI) is a concerted effort from Ministries under the People's Republic of China (PRC) to diplomatically and economically impose its will upon other nations. El Salvador is a US partner and a beneficiary of foreign investment under the BRI. Recent changes to Salvadoran law do not address the implied risks to the nation's supply-chain and cyber infrastructure. This work addresses the gap by exploring previously limited analysis on BRI activities, its intersection with Salvadoran law, and the national security risks introduced by supply-chain reliance from the BRI. This exploratory study examined a portion of the William \& Mary AidData dataset filtered on El Salvador, social media posts, news articles, white papers, and law published by the Legislative Assembly of El Salvador. The analysis suggests that the BRI poses a national security and supply-chain risk to El Salvador through influential-subterfuge, loss of digital sovereignty, which contradicts the State Cybersecurity Agency (ACE) and existing Salvadoran laws. This study provides a foundational understanding and regional context for future research.

\end{abstract}

\bigskip

\section{Introduction} 
Cybersecurity is often referred to as an intersectional community of interest between policy, risk management, and technical controls. The Belt \& Road Initiative represents a strategic cyber security risk from the People's Republic of China (PRC). The Republic of El Salvador (SLV) is assuming cyber infrastructure risk through BRI collaboration with China. Over-reliance on foreign investment allows hostile donors to influence policy and decision-making processes of the recipient country.

Existing research examines BRI as the subject of interest or at a global scale; it fails to address BRI's direct effects on individual countries, especially in Latin America (LATAM) or the forgotten Global South. This gap is more apparent when considering the controversial nature of the topics of statecraft, maligned economic influence, and the notion of impartiality in academia. If a professional discussion is not held, then policy will never change. Given the macroscopic study of an entire region, internal validity around data collection is weakened due to biases and non-random sampling.

The purpose of this paper is to understand the effects of the Belt \& Road Initiative in El Salvador, to characterize the initiative's supply-chain risk to El Salvador's cyber infrastructure, and to compare the initiative's alignment to the Salvadoran national cybersecurity policy.

This paper makes the following contributions: it provides insight into previously underexplored dynamics of the Belt \& Road Initiative in Latin America or \textit{influential-subterfuge}, highlights risks around infrastructure supply-chains \& self-determination or \textit{digital sovereignty}, and discusses the potential conflicts of law and ramifications for El Salvador's national security.

This paper frames the research gap in the cyber-geopolitical environment, qualitative methodology, analysis and findings of research, discussion of their implications, future work, and limitations.

\section{Related Work} 
El Salvador is a nation of approximately 6.5 million citizens \cite{noauthor_population_nodate}. El Salvador has four major Mobile Network Providers (MNPs): Claro, Tigo, Movistar, and Digicel \cite{noauthor_salvador_nodate}. There are at least 20 Internet Service Providers (ISPs) in the country with the majority of IPv4 addresses belonging to Millicom Cable, Telefonica, and ICOMSA \cite{noauthor_top_nodate}.

This problem can be examined as a dichotomy between cybersecurity supply-chain risk and geopolitical policy such as the Belt \& Road Initiative (BRI) of the People's Republic of China (PRC). The BRI is omnipresent across Latin America and Central Asia. El Salvador is one country part of the BRI with very little spotlight.

\subsection{Cyber Supply-Chain Security} Supply-chain security often focuses around critical infrastructure \cite{kimmins_telecommunications_2011}. In 2014, IBM attempted to establish an industry standard for risk management of supply-chains \cite{a_r_szakal_open_2014}. This risk framework was limited to telecommunication providers, but lacks context around the BRI. A literature review in 2021 emphasized that risk is longitudinal and the supply-chain is a popular vector of attack \cite{latif_cyber_2021}. A validity concern with Latif's review derives from a limited search of data for analysis. Machine Learning (ML) and threat intelligence can be applied to predict logistical risk quantitatively \cite{a_yeboah-ofori_cyber_2021}. The use of ML is too narrow in scope and not all cyber problems are easily quantifiable. 

At lower echelons in hardware stacks, Akter examines low-level hardware such as integrated circuits (ICs) \cite{akter_survey_2023}. Akter's research focuses intensively with individual printed circuit boards and microscopic-level analyses. It does not necessarily delve into Chinese components entering vulnerable markets or systems. Ahn examined NIST SP 800-161 and concluded that the current NIST framework is not robust enough to support cyber supply-chain risk management (C-SCRM) for military acquisition programs \cite{ahn_comprehensive_2025}. However, Ahn used a small survey pool of junior personnel to apply use of the Risk Management Framework in a given environment. Future research in this area requiring human subjects needs survey recipients with proper credentials, authorities, and experience levels. 

Sobb concluded that a new model for military supply chains is needed and highlighted the operational risk to national security \cite{sobb_assessment_2020}. For this research, it lacks transferability and it is an example of adding more models to address a problem rather than policy. Luo examined eCrime groups deploying botnets through the Crime-as-a-Service model. In Luo's case, attack tools are tampered or implanted with backdoor functionality unbeknownst to lower-tier eCrime customers or tool operators who unwittingly support botnet amplification \cite{c_luo_your_2025}. Luo's focus was on risk to threat actor tooling origination rather than a risk introduced to a national supply chain. Advanced Persistent Threats are exploiting supply-chain vulnerabilities and affect whole industries \cite{tan_advanced_2025}. Current supply-chain research can be explained through framework evolution, macroscopic policy analyses, infrastructure modeling, or niche hardware analyses. It clearly does not address geopolitical and supply chain risks to national cyber infrastructure.

\subsection{Belt \& Road Initiative}
There is active disagreement in assessing the BRI. It is typically viewed through the lens of economic opportunity. However, Xu coined "globalization-constituting theory" or multi-dimensional perspective of strategic decision based on policy, economics, and social factors \cite{xu_belt_2019}. A significant limitation of Xu's theoretical model derives from the bias and blind trust in Chinese foreign policy. The globalization-constituting theory examines supply-chains at a macroscopic level, but completely ignores the ramifications of the BRI and subtle objectives of strategic cyber adversaries. Conev examined geopolitical friction in Europe, Africa, and Asia in a systematic literature review \cite{conev_modern_2025}. However, Latin America was ignored in the review even though it is a strategic zone of interest. Su found that steel was a leading commodity in exports along the BRI \cite{su_study_2020}. However, Su is only examining a 5-year window from the mid-2010s which does not account for recent shifts in policy or strategy by China. Additionally, Su analyzes trade exports rather than construction, IT infrastructure, or public works projects resulting from BRI modules like the Maritime (MSR) or Digital Silk Road (DSR). Ancillary research indicates that China is actively applying a national surveillance system with machine learning and diffusion of hardware en masse \cite{schuler_computing_2022}. Lin attempted to map Chinese companies to geographic regions in 2023 via a "\textit{difference-in-difference}" framework \cite{lin_investment_2023}. However, Lin only examined the situation from an economic standpoint, not a cyber one.

\subsection{Military-Civil Duality}
China follows a strategy of Military-Civil Fusion in which private industry, military, and government entities collaborate toward a specific objective \cite{bitzinger_chinas_2021}. Public advisories have suspected that Chinese private companies support or engage in malicious cyber activities, often with tacit support of the Chinese government \cite{noauthor_peoples_2024}. Tangentially, the US banned Huawei networking and telecommunications equipment in 2022 due to fears of supply-chain contamination \cite{noauthor_us_2022, lillis_cnn_2022}. The US Department of Defense announced a list of Chinese companies suspected operating congruent with China's Civil-Military Fusion strategy \cite{noauthor_how_nodate, noauthor_us_2022, noauthor_entities_2025}. Based on the People's Liberation Army (PLA) doctrine, China's vision of active defense necessitates creation of \textit{"favorable strategic environments"} and to \textit{"...continuously deepen the close coordination between military conflicts and political, economic, and diplomatic struggles"} \cite{xiao_science_2022}. China's Cybersecurity and Intelligence Laws compel Chinese enterprises to cooperate with their domestic government authorities and process data internally \cite{noauthor_chinas_2026, noauthor_chinas_2023, translate_prc_2017}. Therefore, Chinese companies could be considered a security risk to a nation's telecommunications infrastructure.

\subsection{ORB Networks}
Operational Relay Box (ORB) Networks are a type of hostile cyber architecture for facilitating cyber attacks. There was limited peer-reviewed research around this subject, aside from industry reporting. In 2024, Google (Mandiant) and Team Cymru announced discoveries around suspected Chinese Advanced Persistent Threat (APT) groups using sophisticated traffic proxies to complicate investigations and obfuscate attack origins \cite{raggi_ioc_2024, noauthor_introduction_2024}. Google claims that threat actors typically rely upon multiple cloud virtual private servers (VPS) connected in tandem or compromising residential or weak consumer-grade hardware to build botnets \cite{raggi_ioc_2024}. Research blogs like IntrusionTruth demonstrate a clear connection between cyber attacks and Chinese government contractors or Chinese private sector companies \cite{noauthor_intrusion_2024}. Therefore, Chinese companies are pushing vulnerable hardware, Chinese contractors are complicit in facilitating cyber attacks, and the Chinese government is driving a strategy to maintain a geopolitical advantage against their partner nations.

\subsection{Global Security Initiative}
The Global Security Initiative (GSI) represents an effort to export the Chinese domestic security model to other countries. China is entangling its law enforcement with its partner countries, deploying police as advisors, deploying Chinese AI-powered surveillance equipment in host nations, and providing training \cite{weber_chinas_2025, green_global_2024}. This may appear innocuous, but the purpose of this surveillance is to monitor for dissidents, to target individuals, and to control its populace. In 2023, China's Ministry of Public Security (MPS) had hosted illegal police stations abroad such as in New York City for the purposes of projecting reach and monitoring or harassing expatriates, dissidents, and Chinese-Americans \cite{mozur_invisible_2022, noauthor_two_2023}. With this surveillance model, China has implemented wide-scale smart-city surveillance across its major municipalities and its western regions, employing AI contractors to analyze individuals and collect data including audio, imagery, geo-location, travel, and cellphone telemetry \cite{xiao_video_2022, mozur_invisible_2022}. China has perfected its domestic surveillance. Now, it is exporting that same hardware to other countries. This strategy of domestic security is now in Latin America and helps China impose its will beyond its political borders \cite{egusa_china_2025, fonseca_chinas_2025}.

\subsection{Gap Analysis}
There is very limited cyber research that delves into examining the host-nation effects from supply-chain risk projected through the Chinese Belt \& Road Initiative. Prior research suggests a heavy focus in building frameworks for supply chain risk management, low-level hardware, or strategic supply-chain analysis. Very little research discusses the risk imposed by economic influence in the cyber domain, especially at a geopolitical and strategic level. Now, El Salvador is a prime example of a LATAM nation directly benefiting from and being subverted by the BRI. El Salvador is caught between two peer-competitors (US and China). Therefore, El Salvador is placing itself in an untenable position that degrades its national security in cyberspace.

\subsection{Research Direction}
This qualitative research explores the conceptual model of: "\textit{influential-subterfuge}" from BRI projects and synthesizes possible outcomes, if El Salvador opts not to decouple from the BRI. Secondly, it explores the relationship between China and El Salvador through the latter's cyber policy and the effects of Chinese influence through the BRI.

\section{Methodology}
This study uses an ecological exploratory method. First, it examined the convergence of public policy between China (PRC) and El Salvador (SLV). The intent was furthering examination of the BRI as the intervention and its effects in El Salvador across geopolitical and technical grounds. Second, the study examined Chinese economic and cyber behaviors. Third, Salvadoran law was scrutinized based on its relevance to the BRI.

Data can be described in several categories: public relations (i.e., news), academic (i.e., AidData Project by William \& Mary),  threat intelligence (i.e., industry reporting), and policy (i.e., law). Having multiple data points from varying sources will help in corroboration and triangulation. This will strengthen internal validity of observations \& analyses. By focusing on a LATAM country, transferability is favorable in South America due to competing interests from external powers and regional neutrality.

Academic data represents previous efforts by universities to track the BRI globally. This qualitative information provides context of place, time, and aid type (such as construction, donations, etc.). The AidData corpus was filtered by \textit{"El Salvador"} as a comma-separated value (CSV) document. This table provided the descriptive information and locations for early projects from 2018 to 2023. The file was reviewed through both Claude and manually cross-referenced to ensure relevance. Since BRI projects can be multi-year endeavors or tentative ideas, projects were analyzed based on the year the project was committed (not completed) by China.

Event data was derived from government social media accounts or Salvadoran news outlets. This is active engagement with the general public and highlights high-profile BRI projects which may not be recorded in academic data sets. Between 2024 and the present, BRI projects were identified through X (formerly Twitter) for translated keywords: \textit{"El Salvador"}, \textit{"financial aid"}, \textit{"project"}, or \textit{"relations"}. Tweets were narrowed by relevance using translated keywords: \textit{"computer"}, \textit{"device"}, \textit{"technology"}. These tweets were conveniently selected based on proximity of posters closest to the BRI such as: the account for the Chinese Embassy in El Salvador, accounts for local news organizations, or accounts of private citizens. These tweets were corroborated with secondary resources such as local news articles.

Threat intelligence data originates from well-established cyber-security vendors such as Google/Mandiant or Team Cymru. Threat intelligence reporting provides context around nation-state actors with perceived motivations congruent to national policy and highlights cultural behavior in cyberspace. Analysis of threat intelligence started in reverse chronological order from the present. The latest tactics are at the forefront of concern. The assumption is that longitudinal campaigns require significant investment of resources to prepare the digital battlefield and lesser known targets could become proving grounds for future attacks elsewhere.

Policy data was analyzed in three ways to match the policy and its effects. First-party policy data was acquired from public government websites and translated via Claude AI. Additionally, open-source translated copies of policy data were acquired from third-party think tanks. Foreign text was also corroborated with third-party media reporting and text passed through Google Translate for any text not localized in English.

Non-probabilistic sampling was necessary given that events occurred in the past. Therefore, data was collected both purposely through selection of government sources and conveniently, available without need of travel. This was a deliberate decision based on available resources of time and travel funding.

\section{Analysis \& Results}
\subsection{Geopolitical Complications}
China is engaging in influential-subterfuge to establish strategic dominance over a trade partner nation. These donations, loans, and grant funding provide China the leverage necessary to impose its will upon El Salvador. For example, the BRI gained traction in the country as El Salvador severed diplomatic relations with the Republic of China (Taiwan), which mainland China considers Taiwan a breakaway region \cite{noauthor_statement_2018, noauthor_chinas_2018, noauthor_xi_2019}. Mainland China invested in the El Salvadoran port at La Union which could serve as a logistics hub for the Chinese navy in Latin America \cite{ellis_china_2021-1, malik_banking_2021, jenkins_chinas_2021}. However, the investments are not solely confined to El Salvador (2018); this has been a long-term campaign starting with Costa Rica (2007), Panama (2017), Nicaragua (2021) and Honduras (2023) \cite{pina_china_2024}.

\subsection{AidData Around El Salvador}
Upon reviewing data aggregated through William \& Mary's AidData project, there are approximately 68 events which characterize the suspected BRI-related activity in El Salvador between 2018 and 2023 \cite{noauthor_china_nodate}. One event was excluded because the data collector was unable to independently confirm it starting in 2015 and occurred before the normalization of ties between El Salvador and China in 2017. BRI events were categorized based on perceived funding methods such as donations (financial or material), grant funding, loans, or other (unclear). For this research, data was assessed based on the year which data was committed, not necessarily the year in which the individual project or activity concluded. Although BRI investment is continuous, the highest point of activity occurred in 2020. This occurred shortly after El Salvador severed ties with Taiwan and during the initial phases of the COVID-19 pandemic when global commerce slowed. This was likely a critical moment to invest in El Salvador and to establish a significant presence in the country. Due to social isolation and reduced trade, this would have been the optimal moment to conduct business with little diplomatic resistance.

\begin{table*}[ht]
\centering
\caption{Commitment timeline and funding sources for suspected BRI-related projects. A donation can be financial or in-kind materials (i.e., equipment, food). Note: 1 outlier entry was excluded due to lack of independent confirmation of the transaction.)}
\label{tab:perf_summary}
\begin{tabular}{lcccc}
\hline
\textbf{Year} & \textbf{Donation} & \textbf{Grant} & \textbf{Loan} & \textbf{Other} \\
\hline
2018 & 2 & 2 & 0 & 2 \\
2019 & 9 & 5 & 0 & 2 \\
2020 & 22 & 0 & 3 & 0 \\
2021 & 6 & 0 & 1 & 1 \\
2022 & 3 & 2 & 0 & 0 \\
2023 & 4 & 2 & 0 & 1 \\
\hline
\end{tabular}
\end{table*}

\subsection{Infusion of Chinese Hardware}
Based on the William \& Mary AidData Dashboard, four events intersected with IT infrastructure in El Salvador between 2018 and 2023 \cite{noauthor_china_nodate}. The two most recent events in 2024 and 2026 were not part of the AidData Dashboard.

\begin{itemize}
    \item (2018) El Salvador severed diplomatic ties with Taiwan which enabled grant funding from PRC \cite{noauthor_statement_2018}. Funding was used to procure and distribute 15,000 laptops to 558 schools as part of the Information \& Communications Technology (ICT) Access in Public Educational Centers Project. This may have served as an initial launchpad to test whether or not the partner nation would readily accept hardware donations.
    \item (2020) The PRC Embassy in El Salvador donated 322 Huawei tablets for use by the SLV Ministry of Culture at public sites and venues. Telemetry from these devices could assist Chinese vendors in tracking movement of individuals or groups.
    \item (2021) Under the Trifinio Plan, the PRC provided grant funding for equipment supporting 3 Technology Transfer \& Training Centers between El Salvador, Guatemala, and Honduras. This act could give China the ability to project further influence into neighboring countries or assess the technical skill set of LATAM citizens.
    \item (2023) The PRC Ministry of Finance contributed \$500,000 in grant funding for a submarine telecommunications cable. China's involvement in this sea cable project gives them insight into the peering-point architecture for the country. If this grant money is tied to Chinese equipment, China's own laws compel corporations to serve the interests of the Chinese state first and report vulnerabilities quietly.
    \item (2024) Huawei provided training through the "Seeds of the Future Summit" and hosted the summit in El Salvador \cite{noauthor_huaweis_2024}. The "Seeds of the Future" program allows Huawei to provide training for its 5G cellular technologies, networking hardware, and AI to LATAM students \cite{noauthor_huaweis_2024}. This endeavor helps plant the suggestion for using Chinese hardware and promotes the branding. By using Huawei hardware, students have familiarity with its tech, Chinese equipment is more cost-effective to cash-strapped nations, and the program encourages deployment of Huawei in Latin America. Although no public information can definitively demonstrate Huawei's involvement in espionage, Huawei is a state enterprise beholden to Chinese law. Whatever private information may be known to the US, Australia, and NATO alliance, it was enough to compel them to ban the equipment outright \cite{lu-yueyang_australia_2012, heath_public_2019, kaska_huawei_2019}. Coupled with the Chinese intelligence laws and suspicion of Huawei, wide-scale deployment of Chinese devices introduces possible vulnerabilities into El Salvador's national telecommunications infrastructure.
    \item (2026) The PRC Ministry of Education donated 344,000 laptops and tablets to Salvadoran students \cite{chinese_embassy_in_el_salvador_announcement_2026, slv_ministry_of_education_announcement_2026, ultimahora_retweet_2026}. This laptop donation is another superficial good-will gesture, but it could give the Chinese the ability to track career aspirations or patterns-of-life for Salvadoran minors who may eventually become Salvadoran government employees or military personnel. Unlike the 2018 donation, this 2026 hardware donation represents a significant increase by an approximate factor of 23.
\end{itemize}

\begin{figure}
    \centering
    \includegraphics[width=80mm, scale=0.2]{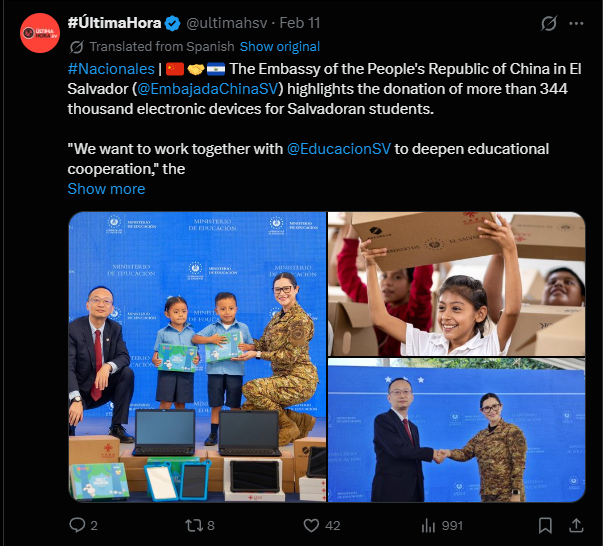}
\caption{Announcement for a donation of 344,000 devices to El Salvadoran students, on February 11, 2026. (Source: Twitter).}
\end{figure}

\subsection{El Salvadoran Cyber-Related Laws}
\textbf{Compliance \& Regulatory Body}.
The \textit{Agencia de Ciberseguridad del Estado (ACE)} or State Cybersecurity Agency was established in November 2024. Its Director-General is an appointee of the President and the organization is given the authority to regulate and enforce compliance for Decree No. 143 (Cybersecurity \& Information Security Law) and Decree No. 144 (Law for Protection of Personal Data) \cite{noauthor_salvador_nodate-1, noauthor_docket_2024, 457_salvador_2025}. It is effectively an equivalent to the US Cyber Infrastructure \& Security Agency (CISA) \cite{noauthor_about_2026}.

\textbf{Cybersecurity \& Information Security Law}. Decree No. 143 is El Salvador's cybersecurity law. One distinction is that this decree affects the public sector exclusively \cite{noauthor_decree_2024-1}. This could be considered an equivalent to the US Federal Information Security Modernization Act (FISMA) \cite{sen_carper_s2521_2014}. This Salvadoran law outlines punitive tiers: minor, serious, and very serious for any delay of response, reporting, or mitigating action in a network environment. Punitive actions can lead to wage-reduction, administrative dismissal, and prohibitions of public employment for up to a decade.

\textbf{Law for Protection of Personal Data}. Decree No. 144 is El Salvador's privacy law. It encompasses privacy protections across both the public and private sectors \cite{noauthor_decree_2024-2}. This Salvadoran law appears to be inspired by the European Union (EU) General Data Protection Regulation (GDPR) \cite{noauthor_general_2016}. It also includes exceptions for credit history or law enforcement investigations.

\textbf{Law for Regulation of Credit History}.
The Salvadoran privacy law does not include credit history because that topic is covered under Decree No. 695 or the \textit{"Law for the Regulation of Information Services on the Credit History of Individuals"} \cite{noauthor_decree_2024, escobar_role_2023}. Decree No. 695 was originally ratified in 2011, but amended several times and most recently in 2025. This is a formal Salvadoran statute as opposed to the US industry standard like Payment Card Industry Data Security Standard (PCI-DSS) \cite{noauthor_pci_2026}.

\section{Discussion}
\subsection{Influential-Subterfuge}
China routinely expresses that the BRI is a mutually-beneficial program between China and its associates. Loans, grants, and donations are a projection of soft-power to influence economically-weaker nations through financial incentives. The Global South or LATAM is often politically neglected. This provided optimal conditions for China to steer countries away from Taiwan. Furthermore, these loans are repaid in money, resources, or property. As literature strongly suggests, these BRI projects carry multiple strategic advantages for China whether or not the recipient country is aware. Chinese companies donating hardware or training keep nations like El Salvador reliant on and subservient to China.

\subsection{Loss of Digital Sovereignty via Supply-Chains}
ACE is a new organization for El Salvador. However, its instantiation is late when compared to the BRI's decade-long influence campaign within the country. The lack of scope to incorporate the private sector places the society at risk. There is a risk that existing Chinese hardware spread across the private sector allows China to potentially extract telemetry on the populace. As previously mentioned, these devices could have questionable security practices or vulnerabilities which enable monitoring of Salvadoran citizens who may become members of the government or military.

Based on the conflicting scopes of El Salvador's recent cyber and privacy laws, ACE simply does not have authority to share threat intelligence with its private sector or support the private sector in securing their own networks. With this legal blind spot, private organizations receiving hardware could find themselves unknowingly part of ORB networks. It is unclear how ACE will support incident response or digital forensic investigations in the private sector.

The BRI provides a vector for China to inject its domestic devices into the target society. This is China engaging in hardware saturation to induce reliance. It leads to a loss of digital sovereignty and self-determination. In extreme hypothetical scenarios, this effectively frames El Salvador's education and IT infrastructure under a house-of-cards.

\subsection{Risk Around El Salvador's Cybersecurity \& Privacy Laws}
As it currently reads, El Salvador's cybersecurity law does not account for private sector hardware. Unlike the US, El Salvador has not banned Chinese hardware. As previously stated, El Salvador accepts Chinese hardware into the private sector. Even more troubling, Decree No. 143, Chapter I, Article 3§J is the Technological-Neutrality clause which upholds \textit{"non-discrimination among technologies"} provided that the technology constitutes \textit{"secure and efficient means"} to comply with the decree or law. This implies that El Salvador will continue to allow Huawei or Chinese hardware in any government capacity as long as the country considers it secure and useful.

Although out of scope for this research, there is concern that the privacy law (Decree No. 144, Chapter I, Article 3§C) gives broad exclusions to: \textit{"public security"}, \textit{"defense"}, \textit{"state security"}, and \textit{"crime prevention"}, while congruent with due process. However, El Salvador is still technically under its State of Exception (Emergency) since 2022, due to its crackdown on gang violence \cite{chang_salvadors_2022, noauthor_salvador_2024, graham_salvador_2025}.

Together, the noticeable gaps within the cybersecurity and data privacy laws expose El Salvador to risk because its government is exempt from data protection laws. This assumes data was processed for national security purposes and that the technology neutrality clause applies. Since Chinese hardware is ubiquitous and cheaper in El Salvador, Chinese devices can process that same data if the state deems Huawei (or another Chinese tech company) a neutral vendor. Thus, the BRI has effectively circumvented the laws intended to protect society.

\section{Future Work}
Future iterations of this research can be improved or expanded through use of other methodologies. Data collection can be refined such as conducting surveys, interviews, or direct field observations in the country. Surveys can be conducted with Salvadoran officials to understand their sentiment or concerns around the BRI or the GSI. Technical action research can be conducted to assess the network health or security for the recipient organizations. Furthermore, the supply-chain risk remains apparent and future research can better support or improve the cybersecurity policy of either government authorities or national telecommunication providers.

\section{Limitations}
This qualitative study yielded several concerns that may affect both internal and external validity.

First, as an exploratory study, there is risk of political bias. There is a sympathetic and cognitive bias from experiencing military work exchanges with allies in El Salvador. As an American cybersecurity practitioner, there is a worry for Chinese APT groups and great-powers competition.

Second, the collection of data is also skewed from non-probabilistic sampling of events over the past several years. The data itself can be contaminated due to social media posts or blog posts by respective governments spinning positive messages. Based on precedent, other qualitative research in this area also suffered internal validity issues stemming from poor sampling techniques or sample sizes. This was an unfortunate experience, but is an acknowledged risk.

Third, AidData is time-boxed between 2017 and 2023. It does not account for more recent and ongoing BRI activity in Latin America. Newer data was decentralized, disparate, or lacked standardized context.

Fourth, external validity issues like transferability may only include Latin American countries or be minimized to El Salvador alone. Nations with stronger economic independence may not be affected or targeted by the BRI. The BRI appears in other countries through varying types of economic projects and not hardware donations or roadworks.

Fifth, there is no immediate or empirical evidence to suggest that the donated devices are infected or unknowingly participating in an ORB network. This is a supply-chain risk being highlighted based on historical questionable business practices or perceived suspicious diplomatic activity.

\section{Conclusion} 
This work explores the complex geopolitical environment between China and El Salvador. It highlights the threat of the BRI through influential-subterfuge, loss of digital sovereignty from hardware saturation, and contradictions in the current cyber \& privacy policies. These findings suggest there are numerous national security and supply-chain risks introduced through BRI participation in El Salvador. This understanding can start a dialogue to improve existing policy in El Salvador. Limitations include non-standardized and non-probabilistic sampling of data and low transferability to countries not economically susceptible.

\printbibliography

@online{noauthor_introduction_2024,
	title = {An Introduction to Operational Relay Box ({ORB}) Networks {\textbar} Team Cymru},
	url = {https://www.team-cymru.com/post/an-introduction-to-operational-relay-box-orb-networks-unpatched-forgotten-and-obscured},
	titleaddon = {Team Cymru},
	publisher = {Team Cymru},
	urldate = {2026-03-26},
	date = {2024-10-29},
	langid = {english},
}

@online{noauthor_two_2023,
	title = {Two Arrested for Operating Illegal Overseas Police Station of the Chinese Government},
	url = {https://www.justice.gov/archives/opa/pr/two-arrested-operating-illegal-overseas-police-station-chinese-government},
	titleaddon = {Department of Justice},
	publisher = {Office of Public Affairs},
	urldate = {2026-04-15},
	date = {2023-04-17},
	langid = {english},
}

@online{fonseca_chinas_2025,
	title = {China’s Quiet Security Push in Latin America},
	url = {https://americasquarterly.org/article/chinas-quiet-security-push/},
	titleaddon = {Americas Quarterly},
	author = {Fonseca, Brian and Writt, Nicole and Brown, Martin},
	urldate = {2026-04-15},
	date = {2025-09-11},
	langid = {american},
}

@article{weber_chinas_2025,
	title = {China's {AI}-Powered Surveillance State},
	volume = {36},
	doi = {10.1353/jod.2025.a970356},
	pages = {151--160},
	number = {4},
	journaltitle = {Journal of Democracy},
	publisher = {Johns Hopkins University Press},
	author = {Weber, Valentin},
	urldate = {2026-04-14},
	date = {2025-10-04},
}

@online{green_global_2024,
	title = {The Global Security Initiative: China’s International Policing Activities},
	url = {https://www.iiss.org/research-paper/2024/10/the-global-security-initiative-chinas-international-policing-activities/},
	shorttitle = {The Global Security Initiative},
	titleaddon = {{IISS}},
	publisher = {International Institute of Strategic Studies},
	author = {Green, Erik and Nouwens, Meia and Nouwens, Veerle},
	urldate = {2026-04-15},
	date = {2024-10-24},
	langid = {english},
}

@online{egusa_china_2025,
	title = {China is Fueling Surveillance Technology Adoption in Latin America—Who is in Charge of Data Privacy?},
	url = {https://securityboulevard.com/2025/09/china-is-fueling-surveillance-technology-adoption-in-latin-america-who-is-in-charge-of-data-privacy/},
	titleaddon = {Security Boulevard},
	author = {Egusa, Conrad},
	urldate = {2026-04-15},
	date = {2025-09-26},
	langid = {american},
}

@article{mozur_invisible_2022,
	title = {‘An Invisible Cage’: How China Is Policing the Future},
	issn = {0362-4331},
	url = {https://www.nytimes.com/2022/06/25/technology/china-surveillance-police.html},
	shorttitle = {‘An Invisible Cage’},
	journaltitle = {The New York Times},
	author = {Mozur, Paul and Xiao, Muyi and Liu, John},
	urldate = {2026-04-15},
	date = {2022-06-26},
	langid = {american},
}

@article{xiao_video_2022,
	title = {Video: China’s Surveillance State Is Growing. These Documents Reveal How.},
	issn = {0362-4331},
	url = {https://www.nytimes.com/video/world/asia/100000008314175/china-government-surveillance-data.html},
	shorttitle = {Video},
	journaltitle = {The New York Times},
	author = {Xiao, Muyi and Mozur, Paul and Qian, Isabelle and Cardia, Alexander},
	urldate = {2026-04-15},
	date = {2022-06-21},
	langid = {american},
}

@online{graham_salvador_2025,
	title = {El Salvador Scraps Term Limits, Paving Way for Bukele to Seek Re-election},
	url = {https://www.bbc.com/news/articles/czd04q87zryo},
	publisher = {{BBC}},
	author = {Graham, Chris},
	urldate = {2026-04-04},
	date = {2025-08-01},
	langid = {british},
}

@online{noauthor_salvador_2024,
	title = {El Salvador: Events of 2024},
	url = {https://www.hrw.org/world-report/2025/country-chapters/el-salvador},
	shorttitle = {El Salvador},
	urldate = {2026-04-04},
	date = {2024-12-08},
	langid = {english},
}

@article{chang_salvadors_2022,
	title = {El Salvador's president has taken over the government and installed martial law},
	url = {https://www.npr.org/2022/09/26/1125172960/el-salvadors-president-has-taken-over-the-government-and-installed-martial-law},
	journaltitle = {{NPR}},
	author = {Chang, Ailsa and Rivera, Enrique and Brown, Ashley},
	urldate = {2026-04-04},
	date = {2022-09-26},
	langid = {english},
}

@online{noauthor_pci_2026,
	title = {{PCI} Security Standards Overview},
	url = {https://www.pcisecuritystandards.org/standards/},
	titleaddon = {{PCI} Security Standards Council},
	urldate = {2026-04-04},
	date = {2026},
	langid = {american},
}

@misc{escobar_role_2023,
	title = {The Role Regulators Play in Closing the Financial Inclusion Gender Gap: A Case Study of El Salvador},
	url = {https://www.afi-global.org/wp-content/uploads/2024/10/El-Salvador_Role-Regulators-Play-in-Closing-the-Financial-Inclusion-Gender-Gap.pdf#:~:text=The%20only%20law%20that%20exists%20is%20the,which%20aims%20to%20ensure%20the%20proper%20handling},
	publisher = {Alliance for Financial Inclusion},
	author = {Escobar, Ana Guadalupe and Umanzor, Cesar Ricardo and Blanco, Clemente Alfredo and Portillo, Ana Idalia and Aquino, Luis and Alfaro, Alejandra and Deras, Daniel},
	urldate = {2026-04-04},
	date = {2023},
}

@online{noauthor_about_2026,
	title = {About {CISA}},
	url = {https://www.cisa.gov/about},
	titleaddon = {Cyber Infrastructure \& Security Agency},
	publisher = {{CISA}},
	urldate = {2026-04-04},
	date = {2026},
	langid = {english},
}

@misc{noauthor_decree_2024,
	title = {Decree No. 695 - Law Regulating Credit History Information Services},
	url = {https://www.asamblea.gob.sv/sites/default/files/documents/decretos/7A4FBD85-7E1B-46BE-9408-6FC549E53E00.pdf},
	publisher = {Legislative Assembly of the Republic of El Salvador},
	urldate = {2026-03-22},
	date = {2024-11-29},
}

@online{noauthor_general_2016,
	title = {General Data Protection Regulation ({GDPR})},
	url = {https://gdpr-info.eu/},
	titleaddon = {General Data Protection Regulation ({GDPR})},
	urldate = {2026-04-04},
	date = {2016},
	langid = {american},
}

@online{noauthor_salvador_nodate,
	title = {El Salvador, June 2022, Mobile Network Experience Report},
	url = {https://insights.opensignal.com/reports/2022/06/elsalvador/mobile-network-experience},
	publisher = {Opensignal},
	urldate = {2026-04-04},
	langid = {english},
}

@online{noauthor_top_nodate,
	title = {Top 20 Internet Service Providers in El Salvador},
	url = {https://db-ip.com/country/SV},
	publisher = {{DB}-{IP}},
	urldate = {2026-04-04},
}

@online{noauthor_population_nodate,
	title = {Population Clock: World},
	url = {https://www.census.gov/popclock/world/es},
	urldate = {2026-04-04},
}

@online{sen_carper_s2521_2014,
	title = {S.2521 - 113th Congress (2013-2014): Federal Information Security Modernization Act of 2014},
	rights = {Text is government work},
	url = {https://www.congress.gov/bill/113th-congress/senate-bill/2521},
	shorttitle = {S.2521 - 113th Congress (2013-2014)},
	type = {legislation},
	author = {Sen. Carper, Thomas R. [D-{DE}},
	urldate = {2026-04-04},
	date = {2014-12-18},
	note = {Archive Location: 2014-06-24},
}

@online{noauthor_statement_2018,
	title = {Statement No. 005 - The R.O.C. Government has Terminated Diplomatic Relations with El Salvador with Immediate Effect in Order to Uphold National Dignity},
	url = {https://en.mofa.gov.tw/https%3a%2f%2fen.mofa.gov.tw%2fNews_Content.aspx%3fn%3d1330%26s%3d34154%26Create%3d1},
	titleaddon = {Press Room Statements \& Responses},
	publisher = {Ministry of Foreign Affairs, Republic of China (Taiwan)},
	urldate = {2026-03-22},
	date = {2018-08-21},
	langid = {english},
}

@report{heath_public_2019,
	title = {Public Evidence of Huawei as a Cyber Threat May Be Elusive, but Restrictions Could Still Be Warranted},
	url = {https://www.rand.org/pubs/commentary/2019/03/public-evidence-of-huawei-as-a-cyber-threat-may-be.html},
	author = {Heath, Timothy R.},
	urldate = {2026-04-04},
	date = {2019-03-07},
	langid = {english},
}

@article{kaska_huawei_2019,
	title = {Huawei, 5G and China as a Security Threat},
	url = {https://ccdcoe.org/uploads/2019/03/CCDCOE-Huawei-2019-03-28-FINAL.pdf},
	pages = {26},
	publisher = {{NATO} Cooperative Cyber Defence Centre of Excellence},
	author = {Kaska, Kadri and Beckvard, Henrik and Minárik, Tomáš},
	date = {2019-03-28},
	langid = {english},
}

@article{lu-yueyang_australia_2012,
	title = {Australia blocks China's Huawei from broadband tender},
	url = {https://www.reuters.com/article/business/australia-blocks-chinas-huawei-from-broadband-tender-idUSBRE82P0GA/},
	journaltitle = {Reuters},
	author = {Lu-{YueYang}, Maggie},
	urldate = {2026-04-04},
	date = {2012-03-26},
	langid = {american},
}

@online{slv_ministry_of_education_announcement_2026,
	title = {Announcement of Donation},
	url = {https://x.com/explore},
	shorttitle = {Ministerio de Educación on X},
	titleaddon = {X (formerly Twitter)},
	author = {{SLV Ministry of Education}},
	urldate = {2026-04-04},
	date = {2026-02-11},
	langid = {english},
}

@online{chinese_embassy_in_el_salvador_announcement_2026,
	title = {Announcement of Donation},
	url = {https://x.com/EmbajadaChinaSV/status/2021682829239730567},
	titleaddon = {X (formerly Twitter)},
	author = {{Chinese Embassy in El Salvador}},
	urldate = {2026-04-04},
	date = {2026-02-11},
	langid = {english},
}

@online{ultimahora_retweet_2026,
	title = {Retweet of Donation},
	url = {https://x.com/ultimahsv/status/2021690569156612444},
	titleaddon = {X (formerly Twitter)},
	author = {{UltimaHora}},
	urldate = {2026-04-04},
	date = {2026-02-11},
	langid = {english},
}

@book{xiao_science_2022,
	title = {The Science of Military Strategy (2020)},
	isbn = {979-8-4050-7762-8},
	pagetotal = {452},
	publisher = {China Aerospace Studies Institute},
	author = {Xiao, Tianliang},
	date = {2022-01},
}

@online{noauthor_intrusion_2024,
	title = {Intrusion Truth},
	url = {https://intrusiontruth.wordpress.com/},
	titleaddon = {Intrusion Truth},
	urldate = {2026-03-29},
	date = {2024-08-07},
	langid = {english},
}

@misc{noauthor_entities_2025,
	title = {Entities Identified as Chinese Military Companies Operating in the United States},
	url = {https://media.defense.gov/2025/Jan/07/2003625471/-1/-1/1/ENTITIES-IDENTIFIED-AS-CHINESE-MILITARY-COMPANIES-OPERATING-IN-THE-UNITED-STATES.PDF},
	publisher = {{US} Department of Defense},
	urldate = {2026-03-29},
	date = {2025},
	langid = {english},
}

@online{noauthor_how_nodate,
	title = {How Military-Civil Fusion Steps Up China's Semiconductor Industry},
	url = {https://digichina.stanford.edu/work/how-military-civil-fusion-helps-chinas-semiconductor-industry-step-up/},
	titleaddon = {{DigiChina}},
	urldate = {2026-03-29},
	langid = {english},
}

@article{bitzinger_chinas_2021,
	location = {Seattle, {WA}},
	title = {China’s Military-Civil Fusion Strategy: Development, Procurement, and Secrecy},
	pages = {1--64},
	number = {1},
	publisher = {National Bureau of Asian Research},
	author = {Bitzinger, Richard A and Evron, Yoram and Yang, Zi},
	date = {2021-01},
	langid = {english},
}

@online{noauthor_us_2022,
	title = {U.S. Restrictions on Huawei Technologies: National Security, Foreign Policy, and Economic Interests},
	rights = {Text is government work},
	url = {https://www.congress.gov/crs-product/R47012},
	shorttitle = {U.S. Restrictions on Huawei Technologies},
	titleaddon = {{US} Congress},
	publisher = {Library of Congress},
	type = {legislation},
	urldate = {2026-03-29},
	date = {2022-01-05},
	langid = {english},
}

@online{lillis_cnn_2022,
	title = {{CNN} Exclusive: {FBI} investigation determined Chinese-made Huawei equipment could disrupt {US} nuclear arsenal communications {\textbar} {CNN} Politics},
	url = {https://www.cnn.com/2022/07/23/politics/fbi-investigation-huawei-china-defense-department-communications-nuclear},
	shorttitle = {{CNN} Exclusive},
	titleaddon = {{CNN}},
	author = {Lillis, Katie Bo},
	urldate = {2026-03-29},
	date = {2022-07-23},
	langid = {english},
}

@online{noauthor_china_nodate,
	location = {Williamsburg, {VA}},
	title = {China Global Development Dashboard},
	url = {https://china.aiddata.org/},
	titleaddon = {{AidData}: A Research Lab by William \& Mary},
	publisher = {William \& Mary},
	urldate = {2026-03-29},
}

@online{noauthor_huaweis_2024,
	title = {Huawei’s Seeds for the Future Summit opens in El Salvador - {InvestinElSalvador}},
	url = {https://web.archive.org/web/20260101152541/https://investinelsalvador.gob.sv/huaweis-seeds-for-the-future-summit-opens-in-el-salvador/},
	urldate = {2026-03-29},
	date = {2024-08-28},
}

@online{translate_prc_2017,
	title = {{PRC} National Intelligence Law (as amended in 2018)},
	url = {https://www.chinalawtranslate.com/national-intelligence-law-of-the-p-r-c-2017/},
	titleaddon = {China Law Translate},
	author = {Translate, China Law},
	urldate = {2026-03-26},
	date = {2017-06-28},
	langid = {english},
}

@online{noauthor_chinas_2026,
	title = {China’s Cybersecurity Law Amendments Increase Penalties, Broaden Extraterritorial Enforcement},
	url = {https://www.lw.com/en/insights/chinas-cybersecurity-law-amendments-increase-penalties-broaden-extraterritorial-enforcement},
	titleaddon = {Latham \& Watkins, {LLP}},
	urldate = {2026-03-26},
	date = {2026-01-14},
	langid = {english},
}

@misc{noauthor_chinas_2023,
	title = {China's National Security Laws: Implications Beyond Borders},
	url = {https://www.cna.org/quick-looks/2023/China-national-security-laws-implications-beyond-borders.pdf},
	publisher = {Center for New America ({CNA})},
	date = {2023-12},
}

@misc{noauthor_peoples_2024,
	title = {People’s Republic of China-Linked Actors Compromise Routers and {IoT} Devices for Botnet Operations},
	url = {https://media.defense.gov/2024/Sep/18/2003547016/-1/-1/0/CSA-PRC-LINKED-ACTORS-BOTNET.PDF},
	publisher = {Federal Bureau of Investigation},
	urldate = {2026-03-26},
	date = {2024-09-18},
}

@online{raggi_ioc_2024,
	title = {{IOC} Extinction? China-Nexus Cyber Espionage Actors Use {ORB} Networks to Raise Cost on Defenders},
	url = {https://cloud.google.com/blog/topics/threat-intelligence/china-nexus-espionage-orb-networks},
	shorttitle = {{IOC} Extinction?},
	titleaddon = {Google Cloud Blog},
	author = {Raggi, Michael},
	urldate = {2026-03-26},
	date = {2024-05-22},
	langid = {english},
}

@misc{noauthor_chinas_2018,
	location = {Washington, {DC}},
	title = {China's Presence in El Salvador and Its Strategies for Gaining Influence},
	url = {https://www.american.edu/centers/latin-american-latino-studies/upload/china-el-salvador-english-final.pdf},
	publisher = {American University},
	urldate = {2026-03-22},
	date = {2018-08-20},
}

@report{malik_banking_2021,
	location = {Williamsburg, {VA}},
	title = {Banking on the Belt and Road: Insights from a New Global Dataset of 13,427 Chinese Development Projects},
	url = {https://docs.aiddata.org/ad4/pdfs/Banking_on_the_Belt_and_Road__Insights_from_a_new_global_dataset_of_13427_Chinese_development_projects.pdf},
	institution = {William \& Mary},
	author = {Malik, Ammar A. and Parks, Bradley and Russell, Brooke and Lin, Joyce Jiahui and Walsh, Katherine and Solomon, Kyra and Zhang, Sheng and Elston, Thai-Binh and Goodman, Seth},
	urldate = {2026-03-22},
	date = {2021-09-21},
}

@misc{noauthor_decree_2024-1,
	title = {Decree No. 143- Law of Cybersecurity \& Information Security},
	url = {https://www.asamblea.gob.sv/sites/default/files/documents/decretos/7A4FBD85-7E1B-46BE-9408-6FC549E53E00.pdf},
	publisher = {Legislative Assembly of the Republic of El Salvador},
	urldate = {2026-03-22},
	date = {2024-11-29},
}

@misc{noauthor_decree_2024-2,
	title = {Decree No. 144 - Law for the Protection of Personal Data},
	url = {https://www.asamblea.gob.sv/sites/default/files/documents/decretos/7A4FBD85-7E1B-46BE-9408-6FC549E53E00.pdf},
	publisher = {Legislative Assembly of the Republic of El Salvador},
	urldate = {2026-03-22},
	date = {2024-11-29},
}

@online{noauthor_xi_2019,
	title = {Xi Jinping Holds Talks with President Nayib Armando Bukele Ortez of El Salvador},
	url = {https://www.fmprc.gov.cn/eng/xw/zyxw/202405/t20240530_11328644.html},
	titleaddon = {Top Stories},
	type = {Ministry of Foreign Affairs - People's Republic of China},
	urldate = {2026-03-22},
	date = {2019-12-03},
	langid = {american},
}

@online{pina_china_2024,
	title = {China Ties Work to Bukele’s Advantage in El Salvador’s Upcoming Election},
	url = {https://thediplomat.com/2024/01/china-ties-work-to-bukeles-advantage-in-el-salvadors-upcoming-election/},
	titleaddon = {The Diplomat},
	author = {Piña, Carlos Eduardo},
	urldate = {2026-03-22},
	date = {2024-01-31},
	langid = {american},
}

@article{jenkins_chinas_2021,
	title = {China’s Belt and Road Initiative in Latin America: What has Changed?},
	volume = {51},
	doi = {10.1177/18681026211047871},
	pages = {13--39},
	number = {1},
	journaltitle = {Journal of Current Chinese Affairs},
	author = {Jenkins, Rhys},
	date = {2021-12},
}

@misc{noauthor_docket_2024,
	location = {San Salvador},
	title = {Docket 186-10-2024-1},
	url = {https://www.asamblea.gob.sv/sites/default/files/documents/dictamenes/06178AAB-C1F9-48B6-959A-423073AE3DBD.pdf},
	publisher = {National Security \& Justice Commission Legislative Palace},
	urldate = {2026-03-22},
	date = {2024-11-11},
}

@inproceedings{sobb_assessment_2020,
	title = {Assessment of Cyber Security Implications of New Technology Integrations into Military Supply Chains},
	doi = {10.1109/SPW50608.2020.00038},
	eventtitle = {2020 {IEEE} Security and Privacy Workshops ({SPW})},
	pages = {128--135},
	booktitle = {2020 {IEEE} Security and Privacy Workshops ({SPW})},
	author = {Sobb, T. M. and Turnbull, B.},
	date = {2020-05-21},
	note = {Journal Abbreviation: 2020 {IEEE} Security and Privacy Workshops ({SPW})},
}

@article{tan_advanced_2025,
	title = {Advanced Persistent Threats Based on Supply Chain Vulnerabilities: Challenges, Solutions, and Future Directions},
	volume = {12},
	issn = {2327-4662},
	doi = {10.1109/JIOT.2025.3528744},
	pages = {6371--6395},
	number = {6},
	journaltitle = {{IEEE} Internet of Things Journal},
	shortjournal = {{IEEE} Internet of Things Journal},
	author = {Tan, Z. and Parambath, S. P. and Anagnostopoulos, C. and Singer, J. and Marnerides, A. K.},
	date = {2025-03-15},
}

@article{akter_survey_2023,
	title = {A Survey on Hardware Security: Current Trends and Challenges},
	volume = {11},
	issn = {2169-3536},
	doi = {10.1109/ACCESS.2023.3288696},
	pages = {77543--77565},
	journaltitle = {{IEEE} Access},
	shortjournal = {{IEEE} Access},
	author = {Akter, S. and Khalil, K. and Bayoumi, M.},
	date = {2023},
}

@inproceedings{kimmins_telecommunications_2011,
	title = {Telecommunications supply chain integrity: Mitigating the supply chain security risks in national public telecommunications infrastructures},
	eventtitle = {2011 Second Worldwide Cybersecurity Summit ({WCS})},
	pages = {1--4},
	booktitle = {2011 Second Worldwide Cybersecurity Summit ({WCS})},
	author = {Kimmins, J.},
	date = {2011-06-01},
	note = {Journal Abbreviation: 2011 Second Worldwide Cybersecurity Summit ({WCS})},
}

@article{ahn_comprehensive_2025,
	title = {Comprehensive Analysis and Recommendation of Supply Chain Risk Management Framework for the Military Domain},
	volume = {13},
	issn = {2169-3536},
	doi = {10.1109/ACCESS.2025.3573515},
	pages = {96813--96833},
	journaltitle = {{IEEE} Access},
	shortjournal = {{IEEE} Access},
	author = {Ahn, J. K. and Cho, K. and Seo, K. and Kim, H. -J. and Kim, S.},
	date = {2025},
}

@article{latif_cyber_2021,
	location = {Poznan},
	title = {Cyber security in supply chain management: A systematic review},
	volume = {17},
	issn = {18952038},
	number = {1},
	journaltitle = {{LogForum}},
	publisher = {Wyzsza Szkola Logistyki},
	author = {Latif, Mohd Nasrulddin Abd and Aziz, Nurul Ashykin Abd and Hussin, Nik Syuhailah Nik and Aziz, Zuraimi Abdul},
	date = {2021},
}

@article{c_luo_your_2025,
	title = {Your Botnet Is His Botnet? A Deep Dive Into the Supply Chain Attack Against Cyber-Arm Industry},
	volume = {33},
	issn = {2998-4157},
	doi = {10.1109/TON.2025.3566903},
	pages = {2319--2335},
	number = {5},
	journaltitle = {{IEEE} Transactions on Networking},
	shortjournal = {{IEEE} Transactions on Networking},
	author = {{C. Luo} and {J. Gan} and {X. Li} and {Z. Pan} and {J. Tang} and {Z. Tian}},
	date = {2025-10},
}

@article{a_yeboah-ofori_cyber_2021,
	title = {Cyber Threat Predictive Analytics for Improving Cyber Supply Chain Security},
	volume = {9},
	issn = {2169-3536},
	doi = {10.1109/ACCESS.2021.3087109},
	pages = {94318--94337},
	journaltitle = {{IEEE} Access},
	shortjournal = {{IEEE} Access},
	author = {{A. Yeboah-Ofori} and {S. Islam} and {S. W. Lee} and {Z. U. Shamszaman} and {K. Muhammad} and {M. Altaf} and {M. S. Al-Rakhami}},
	date = {2021},
}

@article{a_r_szakal_open_2014,
	title = {Open industry standards for mitigating risks to global supply chains},
	volume = {58},
	issn = {0018-8646},
	doi = {10.1147/JRD.2013.2285605},
	pages = {1:1--1:13},
	number = {1},
	journaltitle = {{IBM} Journal of Research and Development},
	shortjournal = {{IBM} Journal of Research and Development},
	author = {{A. R. Szakal} and {K. J. Pearsall}},
	date = {2014-02},
}

@online{noauthor_salvador_nodate-1,
	title = {El Salvador: Cybersecurity and Data Protection Laws Enacted},
	url = {https://www.loc.gov/item/global-legal-monitor/2025-01-21/el-salvador-cybersecurity-and-data-protection-laws-enacted/},
	shorttitle = {El Salvador},
	titleaddon = {Library of Congress, Washington, D.C. 20540 {USA}},
	type = {web page},
	urldate = {2026-02-28},
}

@article{conev_modern_2025,
	location = {Skopje},
	title = {{THE} {MODERN} {CHINESE} {OPENING}-{UP} {POLICY} {VIEWED} {THROUGH} {BRI}: {FROM} {AFRICA} {TO} {EUROPE}},
	volume = {16},
	issn = {18576974},
	pages = {163--170},
	number = {2},
	journaltitle = {{UTMS} Journal of Economics},
	publisher = {University of Tourism and Management},
	author = {Conev, Blagoj and Ilieva, Jana},
	date = {2025-12},
}

@article{xu_belt_2019,
	location = {Bucharest},
	title = {The Belt and Road Initiative and Globalization: The Perspective of Globalization-Constituting Theory},
	volume = {7},
	issn = {23439742},
	pages = {232--237},
	number = {1},
	journaltitle = {Global Economic Observer},
	publisher = {Nicolae Titulescu University Editorial House},
	author = {Xu, Fayin},
	date = {2019},
}

@online{457_salvador_2025,
	title = {El Salvador {ICT} Cybersecurity and Information Security Law},
	url = {https://www.trade.gov/market-intelligence/el-salvador-ict-cybersecurity-and-information-security-law},
	author = {457},
	urldate = {2026-02-16},
	date = {2025-01-24},
	langid = {english},
}

@article{schuler_computing_2022,
	location = {New York, {NY}, {USA}},
	title = {Computing as Oppression: Authoritarian Technology Poses a Worldwide Threat},
	volume = {3},
	url = {https://doi.org/10.1145/3568400},
	doi = {10.1145/3568400},
	number = {4},
	journaltitle = {Digit. Gov.: Res. Pract.},
	publisher = {Association for Computing Machinery},
	author = {Schuler, Douglas},
	date = {2022-12},
}

@inproceedings{su_study_2020,
	location = {New York, {NY}, {USA}},
	title = {Study on the Impact of the Level of Trade Facilitation of Countries along the Belt and Road Initiative on China's Steel Products Export},
	isbn = {978-1-4503-6248-1},
	url = {https://doi.org/10.1145/3377672.3378059},
	doi = {10.1145/3377672.3378059},
	series = {{AMME} 2019},
	pages = {206--209},
	booktitle = {Proceedings of the 2019 Annual Meeting on Management Engineering},
	publisher = {Association for Computing Machinery},
	author = {Su, Qiaoqin and Gu, Jianqiang},
	date = {2020},
}

@article{lin_investment_2023,
	location = {Hershey},
	title = {Investment Efficiency of Chinese Enterprises Under ‘Belt and Road' Initiatives: What Information Do We Get From Studies?},
	volume = {31},
	issn = {10627375},
	doi = {10.4018/JGIM.323360},
	pages = {1--20},
	number = {1},
	journaltitle = {Journal of Global Information Management},
	publisher = {{IGI} Global},
	author = {Lin, Boqiang and Lin, Ping and Lin, Mengting},
	date = {2023},
}

@online{ellis_china_2021-1,
	title = {China and El Salvador: An Update},
	url = {https://www.csis.org/analysis/china-and-el-salvador-update},
	shorttitle = {China and El Salvador},
	publisher = {Center for Strategic \& International Studies},
	author = {Ellis, Evan},
	urldate = {2026-02-16},
	date = {2021-03-22},
	langid = {english},
}

\end{document}